\documentclass[pr,nofootinbib,twocolumn,showpacs,aps]{revtex4-1}
\usepackage{bm}
\usepackage{graphicx}
\usepackage{amssymb}
\usepackage{amsmath}
\usepackage{eufrak}
\usepackage{color}
\usepackage[utf8]{inputenc}
\usepackage[unicode=true,colorlinks=true,urlcolor=blue,citecolor=blue]{hyperref}

\usepackage{pifont}
\usepackage{ulem}

\begin{document}

\title{Bulk inversion asymmetry induced magnetogyrotropic reflection from quantum wells}

\author{L.\,V.\, Kotova$^{1}$}
\author{V. N. Kats$^2$} 
\author{A. V. Platonov$^2$} 
\author{V. P. Kochereshko$^2$}
\author{R. Andr\'e$^3$}  
\author{L. E. Golub$^2$}
\affiliation{$^1$ITMO University, 197101 St.~Petersburg, Russia} 
\affiliation{$^2$Ioffe Institute, 194021 St.~Petersburg, Russia}
\affiliation{$^3$Universit\'e Grenoble Alpes, CNRS, Institut NEEL, F-38000 Grenoble, France}

\begin{abstract}
Bulk inversion asymmetry (BIA) of  III-V and II-VI semiconductor quantum wells is demonstrated by reflection experiments in magnetic field oriented in the structure plane. The linear in the magnetic field 
contribution to the reflection coefficients is measured at oblique incidence of $s$ and $p$ polarized light in vicinity of exciton resonances. 
We demonstrate that this contribution to the reflection is caused by magnetogyrotropy of quantum wells, i.e. by the terms in the optical response which are linear in both the magnetic field strength and light wavevector.
Theory of magnetogyrotropic effects in light reflection is developed with account for linear in momentum BIA induced terms in the electron and hole effective Hamiltonians. Theoretical estimates agree with the experimental findings. We have found the electron BIA splitting constant in both GaAs and CdTe based quantum wells is about three times smaller than that for heavy holes.
\end{abstract}

\maketitle{}

\section{Introduction}

In low-symmetry semiconductor quantum wells (QWs), various remarkable effects are present which are absent in bulk semiconductors and symmetric structures. The examples are spin-dependent phenomena, e.g. electrical spin orientation, conversion of nonequilibrium spin into electric current, electric-field induced spin rotations, as well as nonlinear optical effects like 
photogalvanics, for a review see Ref.~\cite{LG_SG_review}. These effects are symmetry-allowed in gyrotropic systems where some components of vectors (e.g. electric current) and pseudovectors (spin, magnetic field) transform identically under point symmetry operations~\cite{NonRecipReview}. The sources of gyrotropy are the bulk and the structure inversion asymmetries present in most of QW structures grown from III-V or II-VI semiconductors. The Structure Inversion Asymmetry (SIA) is present if the QW has different barrier materials or an  electric field removes symmetry in the growth direction. Bulk Inversion Asymmetry (BIA) is present even in QWs with symmetric heteropotential. BIA is caused by an absence of inversion center in the bulk material forming the QW layer. The most popular (001) QWs have the point symmetry group D$_{2d}$ which is gyrotropic, and the BIA-induced effects are studied intensively~\cite{Culcer-Hamilton-Winkler}.

Gyrotropy  manifests itself in optics, and a powerful tool for its study is light reflection  experiments. In particular, gyrotropy results in conversion of light polarization state at reflection. In QWs, an equivalence of the in-plane components of the photon momentum to an effective magnetic field result in natural optical activity~\cite{Opt_act_PRB}. 
The new group of phenomena called {\it magnetogyrotropic} effects take place in the presence of an external magnetic field~\cite{MSD_thin films,magn_gyr_review}. 
Magnetogyrotropy means the terms in the optical response which are linear in both the magnetic field strength and the photon wavevector.
The magnetogyrotropy is absent in centrosymmetric systems, and in QWs it also stems from inversion asymmetry.
The SIA induced magnetogyrotropy has been demonstrated recently  in reflection experiments~\cite{MSD_PRB}. 
The combination of SIA and magnetic field result in interesting phenomena in  light emission which
are greatly enhanced 
in semimagnetic QWs with grating~\cite{Spitzer}. 
However, for the study of these effects, special asymmetric design of QWs and hybrid plasmonic structures has been used in Refs.~\cite{MSD_PRB,Spitzer}. By contrast, the BIA-induced magnetogyrotropy does not require special technological efforts. In the present work we use the fact that any QW grown from GaAs or CdTe is gyrotropic due to BIA, and demonstrate the BIA-induced magnetogyrotropy effects. 
We investigate reflection from QWs in vicinity of exciton resonances where the magnetogyrotropic effects are greatly enhanced. 

Symmetry analysis allows us to 
choose a proper orientation of magnetic field and light incidence plane for study of the BIA contribution to light reflection. 
For this purpose we find the BIA
contribution 
bilinear in both the photon wavevector $\bm q$ and magnetic field $\bm B$ to the nonlocal dielectric susceptibility tensor $\hat{\bm \chi}$.
For (001) QWs
(point symmetry group $D_{2d}$),
symmetry analysis and the Onsager principle yield the following magnetogyrotropic contributions to the susceptibility
\begin{align}
\label{phenom}
\chi_{xx}\pm\chi_{yy} = T_{\pm} (q_xB_x \mp q_yB_y), \nonumber \\
\chi_{xy}=\chi_{yx} = T (q_yB_x-q_xB_y).
\end{align}
Here $x,y$ are $\langle 100 \rangle$ axes in the QW plane, and $T$, $T_+, T_-$ are three linearly-independent functions which will be found below.

In experiments, magnetogyrotropy manifests itself as an additional birefringence caused by both the magnetic field $\bm B$ and light wavevector $\bm q$. 
These magnetogyrotropic contributions can be probed in reflection experiments.
By contrast, the pure $\bm B$-linear terms in the reflection are forbidden by the time-inversion symmetry. 
It follows from Eq.~\eqref{phenom}
that the reflection coefficients at oblique incidence acquire the contributions linear in both $\bm q$ and $\bm B$ 
for $s$ and $p$ polarized light
\begin{equation}
\label{r_phenom}
\Delta r \propto q_\parallel B_\parallel,
\end{equation}
where symbol $\parallel$ denotes projections onto the QW plane.
This relation holds for 
incidence plane oriented 
along $\langle 100 \rangle$
crystallographic directions. Equation~\eqref{r_phenom} demonstrates that in the $\bm B$-linear contribution to the reflection coefficient  is present due to BIA if the magnetic field lies in the light incidence plane. We note that the contribution caused by SIA is absent in this geometry. For its observation the magnetic field should be oriented perpendicular to the incidence plane~\cite{MSD_PRB}. This allows us to study pure BIA magnetogyrotropic effect by choosing the geometry  $\bm B \parallel \bm q_\parallel$.

The paper is organized as follows. In Sec.~\ref{exp} we describe our experiments and deduce the size of the magnetogyrotropy signal for studied QWs. In Sec.~\ref{theory},
the microscopic theory accounting for BIA terms in the electron and hole effective Hamiltonians is developed allowing for finding the magnetoinduced correction to the reflection coefficient. In Sec.~\ref{Disc} we compare experimental results with theory and estimate the electron and hole BIA spin-splittings. Concluding remarks are given in Sec.~\ref{Concl}.

\section{Experiment}
\label{exp}

\begin{figure}[b]
\includegraphics[width=0.5\linewidth]{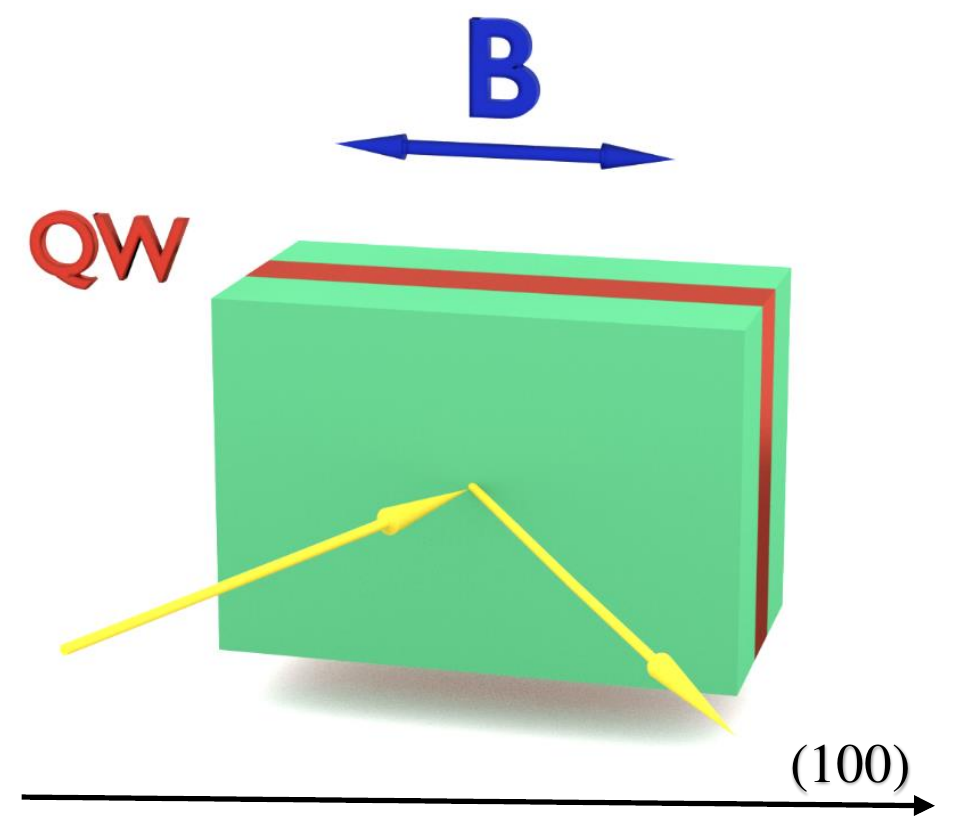}
\includegraphics[width=0.4\linewidth]{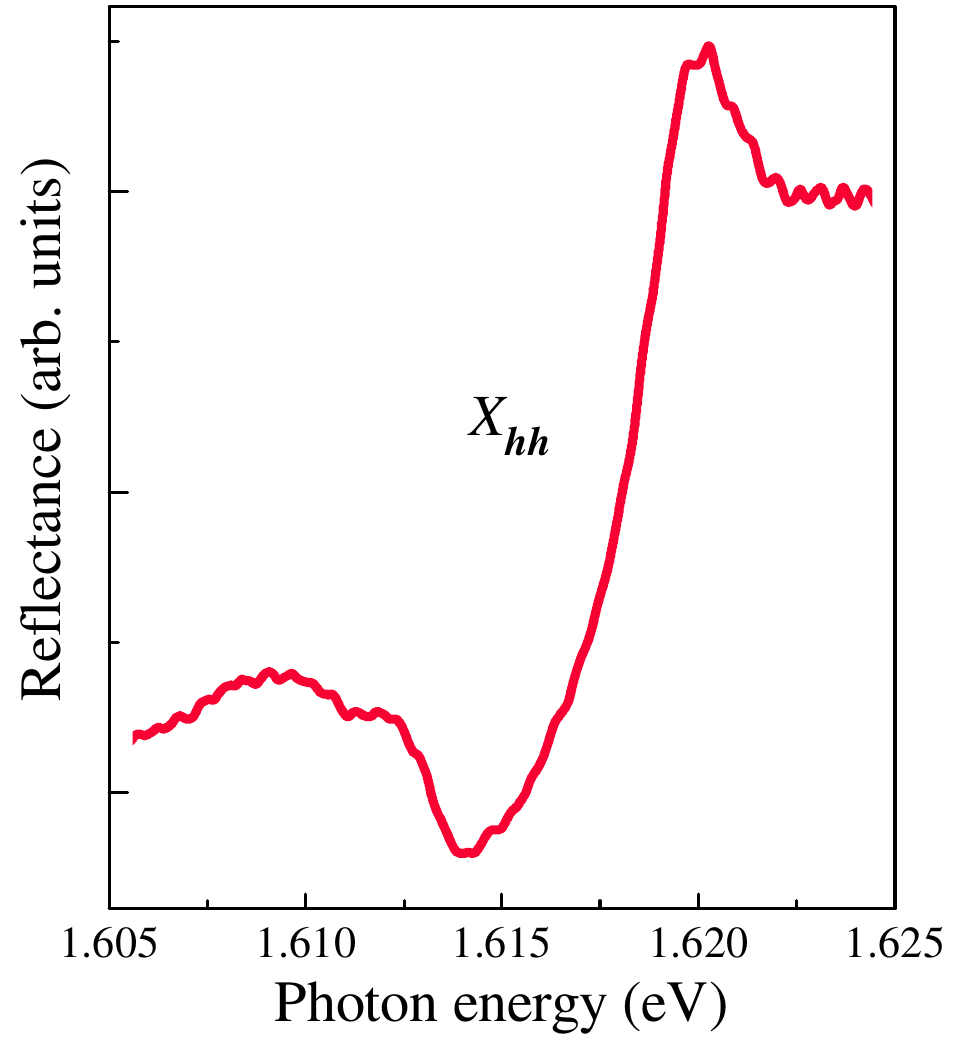}
\caption{Left: Experimental geometry. The magnetic field $\bm B$ lies the plane of the QW and the light incidence plane contains $\bm B$. 
Right: Reflectance spectrum measured from CdTe asymmetric heterostructure in the vicinity of the $X_{hh}$ resonance. 
}
\label{fig:Reflectance}
\end{figure}

For experimental investigation of BIA-induced magnetogyrotropic effects we studied   CdTe- and GaAs based samples with single (001) QWs, Fig.~\ref{fig:Reflectance}(a).
A triangular GaAs/AlGaAs QW was grown by the molecular beam epitaxy method on a semi-insulating substrate. The structure contains a 200~nm wide Al$_{0.28}$Ga$_{0.72}$As barrier followed by the 8~nm wide QW. Then the other sloping barrier was grown 
with Al concentration smoothly increasing from 4~\% to 28~\% on a layer of width 27~nm. 
The rectangular 8~nm wide Cd$_{0.9}$Zn$_{0.1}$Te/CdTe/Cd$_{0.4}$Mg$_{0.6}$Te QW with 90~nm wide barriers was grown on a buffer layer to a Cd$_{0.96}$Zn$_{0.04}$Te substrate. 
The  design of both structures is identical to that of the samples used in Ref.~\cite{MSD_PRB}.

Magnetic field was oriented in the QW plane.
The magnetic field up to 1~T was produced by the electromagnet with a ferromagnetic core.
Experiment was performed at temperature $T=3$~K in closed cycle helium cryostat which was located in the core gap. The geometry of electromagnet and cryostat allowed for oblique light incidence.
We measured polarization of the  reflected light 
at oblique incidence with the incidence angle $\theta_0=27^\circ$.
A halogen lamp was used as a light source for reflection measurements. Lenses and slits formed parallel light beam. Glan-Taylor prisms produced linearly polarization. Excitation was linearly polarized in the plane of incidence ($p$ polarization) and perpendicular to the incidence plane ($s$ polarization).
Spectral dependencies of the reflected light were registered by a CCD camera conjoined with a monochromator. Strong exciton resonances are present in experimental data for both samples.
As an example, the reflection spectrum is shown in Fig.~\ref{fig:Reflectance}(b) for CdTe QW structure. The heavy-hole exciton resonance  $X_{hh}$ is clearly seen.

\begin{figure}[t]
\includegraphics[width=0.6\linewidth]{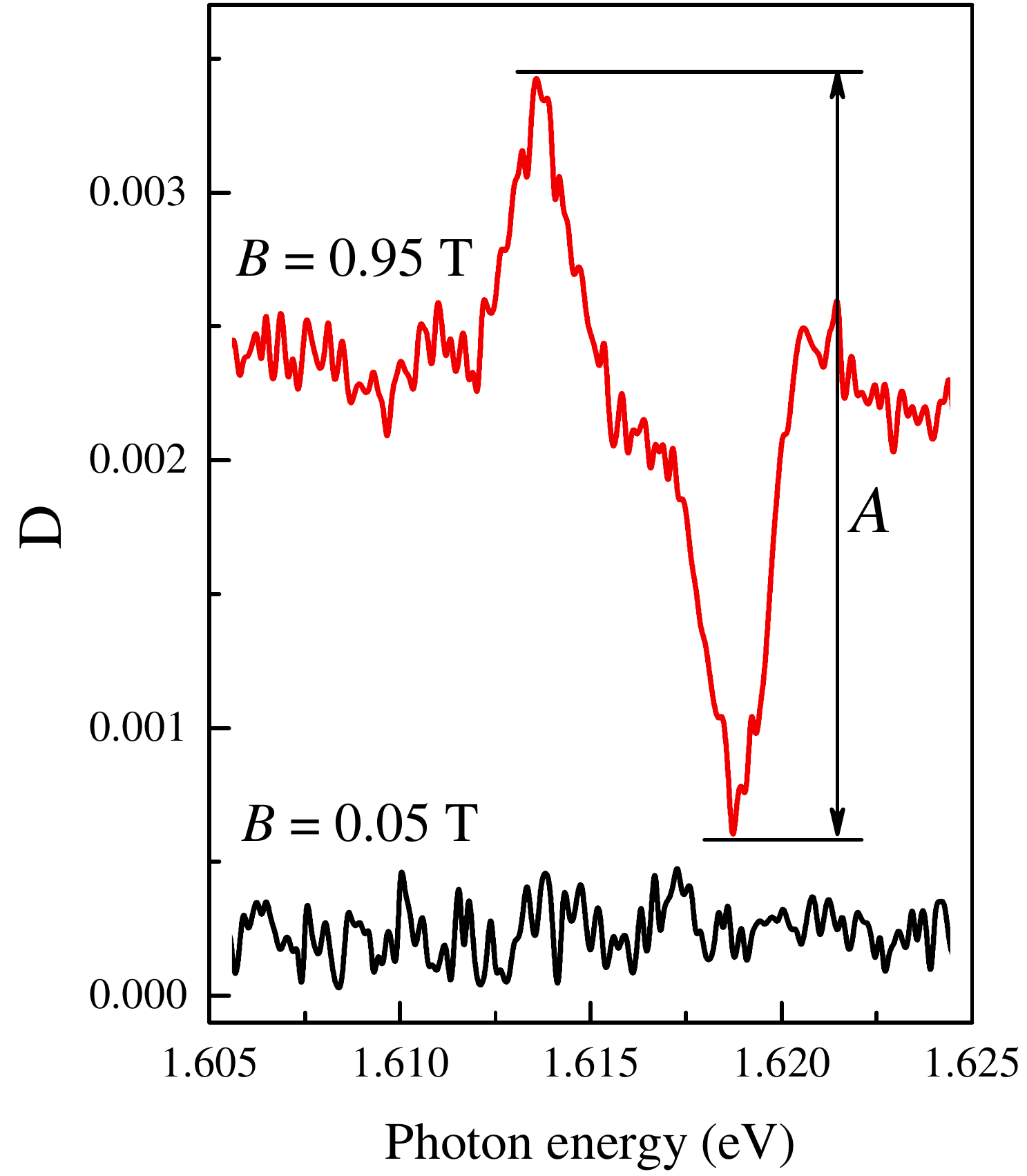}
\caption{Signal $D$, Eq.~\eqref{D}, for $p$ polarized incident light reflected from CdTe sample in the vicinity of the $X_{hh}$ resonance at magnetic fields $B=0.05$~T and $B=0.95$~T. The arrow indicates the signal amplitude $A$ plotted in Fig.~\ref{fig:Amplitudes_CdTe}.
}
\label{fig:MSD_spectrum_CdTe}
\end{figure}

\begin{figure}[h]
\includegraphics[width=0.7\linewidth]{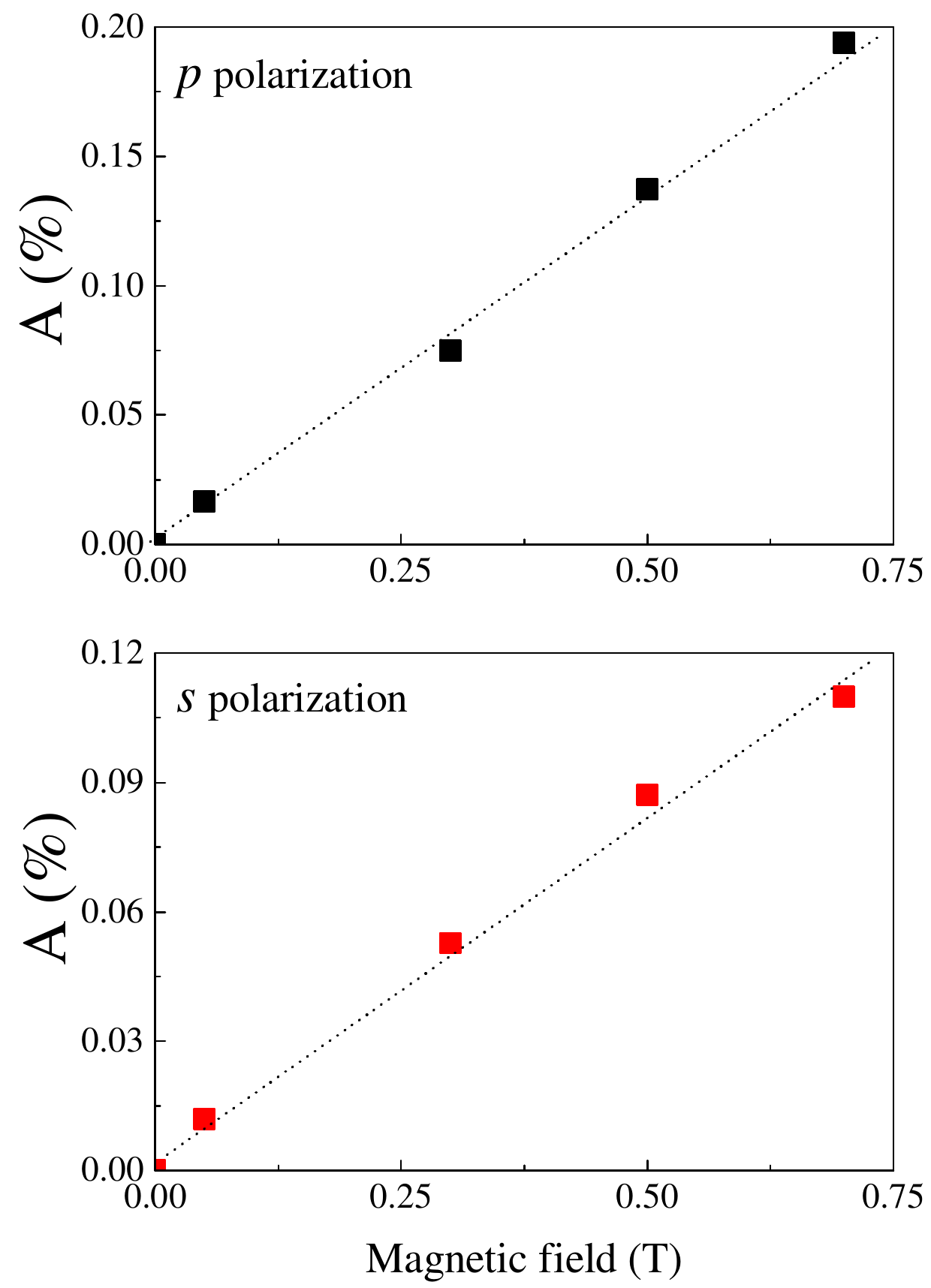}
\caption{Magnetic field dependencies of the signal amplitudes defined in Fig.~\ref{fig:MSD_spectrum_CdTe} for $s$ and $p$ polarized incident light for the CdTe sample. The lines are guides for eyes.
}
\label{fig:Amplitudes_CdTe}
\end{figure}

We measured the  polarization components of the reflected light in magnetic fields from $-1$~T to $+1$~T.
We used both $s$ and $p$ polarized light and detected the 
reflection coefficients $r_{s,p}$.
We analyzed the odd in $\bm B$ contribution to reflection:
\begin{equation}
\label{D}
D = {r(\bm B) - r(-\bm B)\over r(\bm B) + r(-\bm B)}.
\end{equation}
The spectra $D(\omega)$ for  CdTe QW are plotted in Fig.~\ref{fig:MSD_spectrum_CdTe} for $p$ polarization.
In order to quantify the effect of magnetic field we determine the amplitude $A$ from each reflection spectrum. 
The dependencies of the amplitudes on the magnetic field strength are shown in Fig.~\ref{fig:Amplitudes_CdTe}. The dependencies $A(B)$ are linear up to $B=0.75$~T. The amplitude  is larger for $p$ polarization.

\begin{figure}[t]
\includegraphics[width=0.8\linewidth]{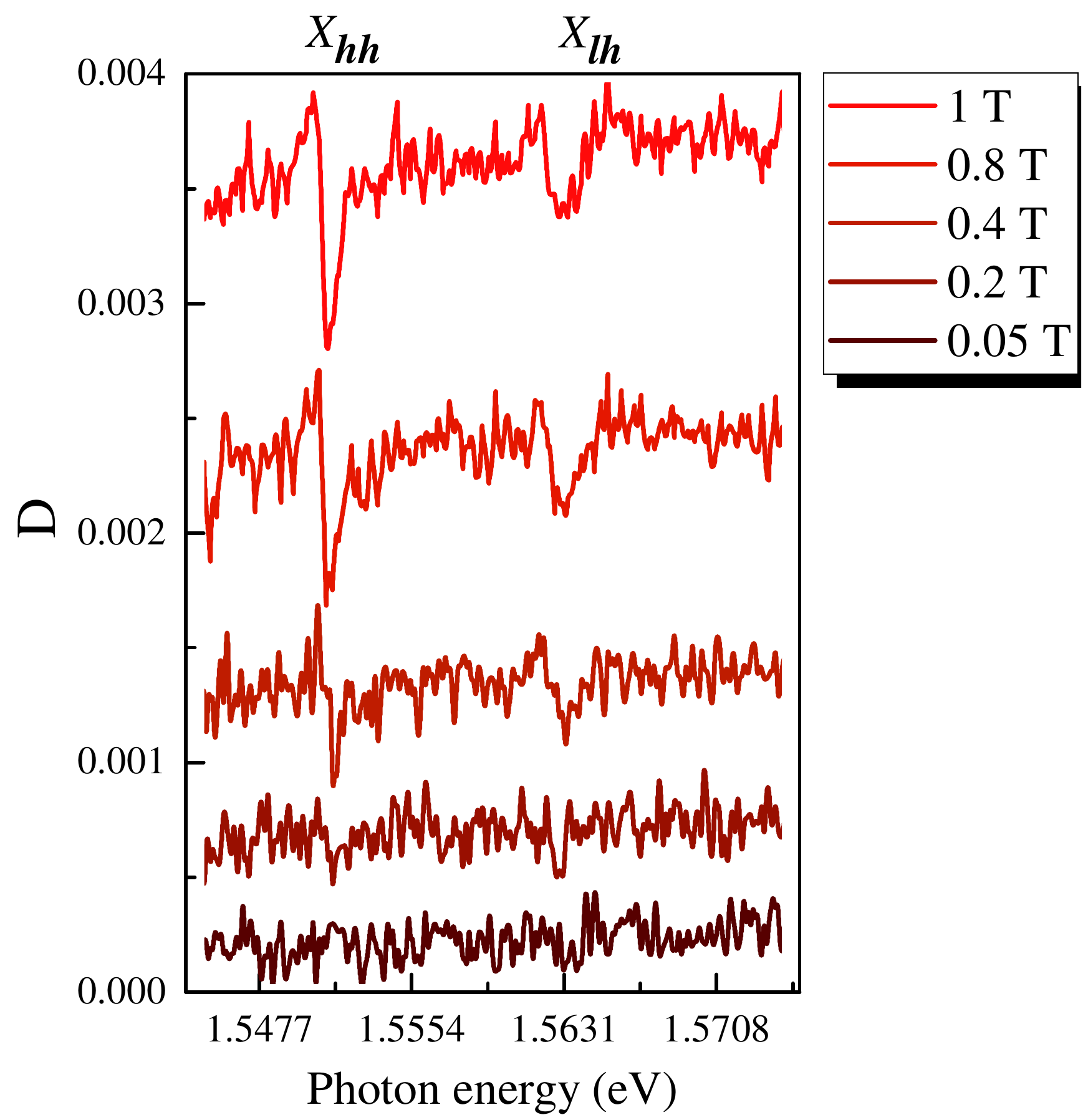}
\caption{The signal for $p$ polarized incident light reflected from GaAs/AlGaAs asymmetric heterostructure at $T=3$~K.  The heavy-hole and light-hole exciton resonances are indicated. 
}
\label{fig:Waterfall_GaAs}
\end{figure}

\begin{figure}[h]
\includegraphics[width=0.7\linewidth]{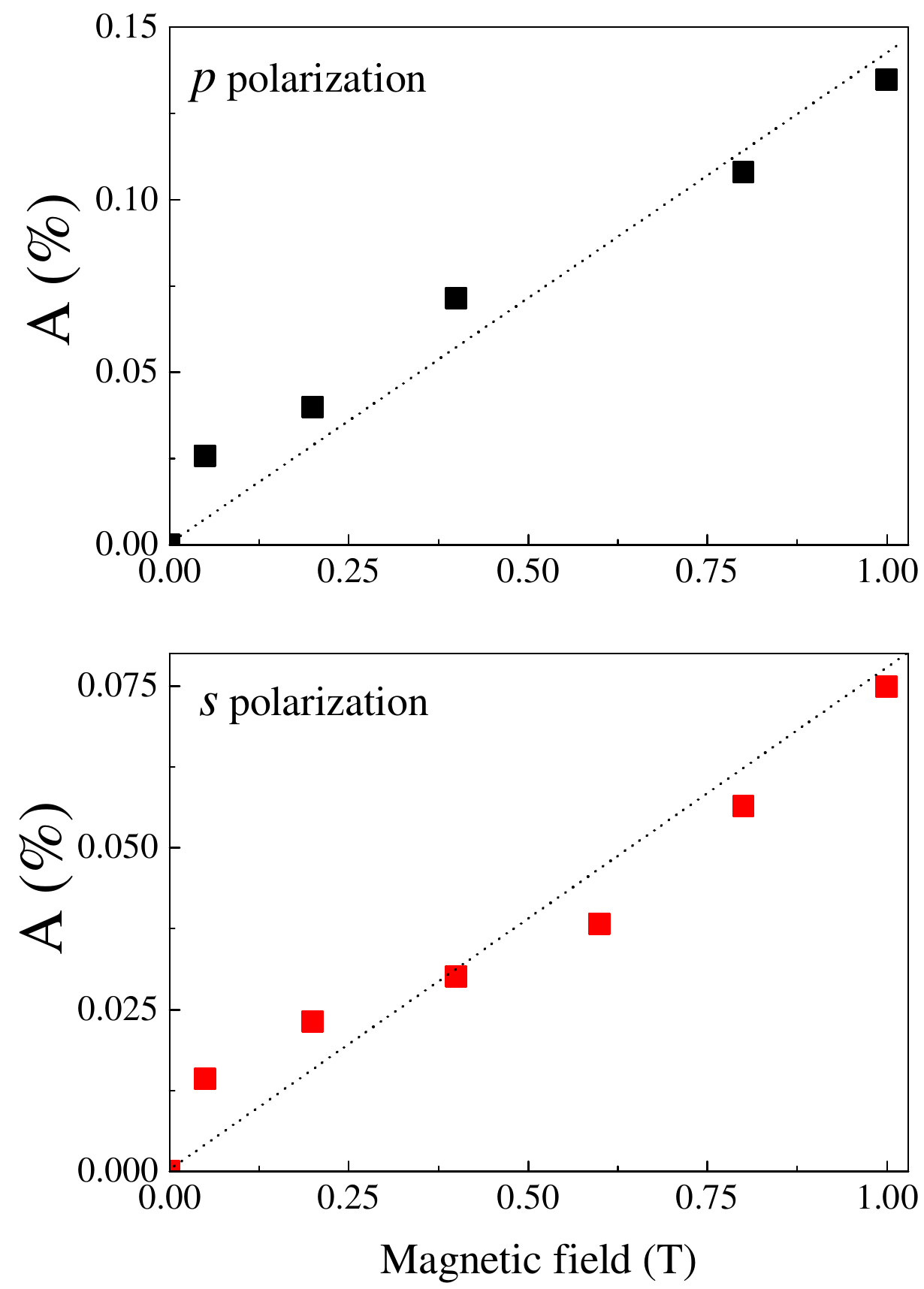}
\caption{Magnetic field dependencies of the signal amplitudes for the GaAs/AlGaAs sample. The amplitudes are defined as in Fig.~\ref{fig:MSD_spectrum_CdTe} 
for $s$- and $p$ polarized  light. 
}
\label{fig:Amplitudes_GaAs}
\end{figure}

We performed the same studies on the sample with the GaAs QW. The spectra $D(\omega)$ for five values of the magnetic field are shown in Fig.~\ref{fig:Waterfall_GaAs}. Increase of the signal amplitude is clearly seen. The dependence of the amplitude $A$ on the magnetic field is shown in Fig.~\ref{fig:Amplitudes_GaAs} for both $s$- and $p$ polarization of incident light. 

To summarize the experimental part, the amplitudes of the $B$-linear contribution to the reflection coefficients in the CdTe QWs are equal to $2.8 \times 10^{-3} B$~T$^{-1}$ and $1.5 \times 10^{-3} B$~T$^{-1}$  for $p$ and $s$ polarization, respectively. For GaAs structure, the amplitudes are $1.5 \times 10^{-3} B$~T$^{-1}$ and $0.75 \times 10^{-3} B$~T$^{-1}$.


%

\section{Theory}
\label{theory}

With account for the nondiagonal terms of the Luttinger effective Hamiltonian~\cite{EL_book}, the heavy-hole wavefunctions in QWs at $B=0$ have the form
\begin{equation}
	\Psi_{hh1, \pm 3/2}^{B=0} =  \phi_{hh1}(z) u_{\pm 3/2}
	\mp  {\gamma_3\hbar^2 \over m_0}  \sum_n { \{k_zk_\pm\} \phi_{lhn}(z) \over E_{hh1}-E_{lhn}}  u_{\pm 1/2}.
\end{equation}
Here $\bm k$ is the hole wavevector,  $k_\pm = k_x \pm ik_y$, $\{\ldots\}$ denotes the anticommutator, $\phi_{hh1}$ and
$\phi_{lhn}$ are the functions of size quantization of corresponding levels of heavy and light holes, $E_{hh1}$ and $E_{lhn}$ are their energies,
$\gamma_3$ is the Luttinger parameter, 
and $u_{\mu}$ are the Bloch amplitudes for the states of the top of the valence band.

In magnetic field $\bm B \perp z$, we choose the vector potential $\bm A=z(B_y,-B_x,0)$, and the Peierls substitution yields $k_\pm \to k_\pm \mp izB_\pm e/(\hbar c)$.
Therefore we obtain the $\bm B_\parallel$-dependent wavefunction
\begin{equation}
\label{Psi_B}
\Psi_{hh1, \pm 3/2}^{\bm B} =  \phi_{hh1}(z) u_{\pm 3/2}
	+ B_\pm F(z)u_{\pm 1/2}\equiv \left|h,\mp 1/2\right>,
\end{equation}
where
\begin{equation}
F(z) = {\gamma_3 e\hbar \over m_0 c} \sum_n { \{i k_z z\} \phi_{lhn}(z) \over E_{hh1}-E_{lhn}}.
\end{equation}

The matrix elements of exciton creation are ${\bm e\cdot \bm d_{e,s;h,m}= \langle e,s |\bm e\cdot \hat{\bm d}|{\cal K}(h,m)\rangle}$ where $\bm e$ is the light polarization vector, $\hat{\bm d}$ is the dipole momentum operator, and ${\cal K}=i\sigma_y{\cal K}_0$ is the time inversion operator with ${\cal K}_0$ being the complex conjugation operation.
Here $s=\pm 1/2$ and $m=\pm 1/2$ enumerates the spin-degenerate states in the conduction and valence bands, respectively.
As a result of the magnetic field induced mixing of the hole states, Eq.~\eqref{Psi_B}, all four optical transitions become allowed. The matrix elements of the components of the dipole momentum operator $\bm e\cdot \bm d_{e,s;h,m}$ in the basis $e,\pm 1/2$, $h,\pm 1/2$ are given by
\begin{equation}
\label{ed}
\bm e\cdot \bm d= 
d_\perp \left( 
\begin{array}{cc}
e_+ + 2\zeta B_+ e_z & -e_- \zeta B_- \\
-e_+ \zeta B_+ & e_- - 2\zeta B_+ e_z
\end{array}
\right).
\end{equation}
Here $d_\perp = iep_{cv}\left<e1 |hh1\right>/(m_0 \omega_0\sqrt{2})$, $p_{cv}$ and $\omega_0$ are the interband momentum matrix element and the exciton frequency, respectively,
and the parameter $\zeta$ is given by
\begin{equation}
\zeta = {\left<e1 |F(z)\right> \over \left<e1 |hh1\right>}
= {\gamma_3e\hbar\over m_0 c \sqrt{3} \left<e1 |hh1\right>} \sum_n {\left< e1| \{i k_z z\} |lhn\right> \over E_{hh1}-E_{lhn}}.
\end{equation}
Here summation is performed over the light-hole states with even envelopes ($n=1,3,5\ldots$).

The nonlocal exciton dielectric polarization $\bm P$ in quantum wells depends on the growth-direction coordinate $z$ 
\begin{equation}
\bm P(\bm q, z) = \Phi(z)\sum_{\nu} \bm d_\nu^*(\bm q) C_\nu(\bm q).
\end{equation}
Here summation is performed over four exciton states $\nu=(e,s;h,m)$, $s,m=\pm1/2$,
and $\Phi(z)$ is the envelope function of the exciton size quantization at coinciding coordinates of electron
and hole~\cite{EL_book}.
The coefficients $C_\nu$ satisfy the equation
\begin{multline}
\left[ (\hbar\omega_0 - \hbar\omega - i\Gamma)\delta_{\nu \nu'}+{\cal H}_{\nu \nu'}(\bm q) \right] C_{\nu'} \\ = \int dz' \Phi^*(z')\bm E(z') \cdot \bm d_\nu(\bm q)  .
\end{multline}
Here $\omega_0$ and  $\Gamma$ are the heavy-hole resonant frequency and
linewidth, $\hat{\cal H}$ is the contribution to the exciton Hamiltonian caused by the spin-orbit interaction,
and $\bm E(z)$ is the total electric field. 
The first order in the spin-orbit interaction correction to the polarization describing the magnetogyrotropy is given by
\begin{equation}
\label{dP}
\delta \bm P(\bm q, z) = - {\Lambda_0  \Phi(z)\over (\hbar\omega_0 - \hbar\omega - i\Gamma)^2} \sum_{\nu\nu'} \bm d_\nu^*  {\cal H}_{\nu\nu'} (\bm d_{\nu'} \cdot \bm E_0).
\end{equation}
Here $\Lambda_0 = \int dz \Phi^*(z) \exp{(iq_zz)}$, and we neglect radiative renormalizations of 
$\omega_0$ and $\Gamma$.

We account for BIA via the $\bm k$-linear spin-orbit splitting of the conduction- and valence band states. As a result the total spin-orbit Hamiltonian is a sum of the electron and hole terms which have the following forms in the basis $e1 \uparrow$, $e1 \downarrow$ of the electron states and $\left|h,1/2\right>$, $\left|h,-1/2\right>$ of the hole states~\cite{High_Butov_Kavokin,MD_MG_EL_PRB_2014}: 
\begin{equation}
\label{H_SO}
{\cal H} = {\cal H}^e+{\cal H}^h,
\qquad
{\cal H}^{e,h} = \beta_{e,h}(\sigma_x^{e,h} k_x^{e,h} - \sigma_y^{e,h} k_y^{e,h}) .
\end{equation}
Here $x\parallel [100]$, $y\parallel [010]$, $\bm k^{e,h}$ are the electron and hole wavevectors in the QW plane, and $\beta_{e,h}$ are the two-dimensional Dresselhaus constants. 
Using Eqs.~\eqref{ed},~\eqref{dP} and~\eqref{H_SO} we obtain the magnetogyrotropic contributions to the nonlocal susceptibility $\hat{\bm \chi}$
defined as $\delta \bm P(z) = \int dz' \hat{\bm \chi}(z,z') \bm E(z')$
in the form of Eqs.~\eqref{phenom} with
\begin{equation}
T_+= \tilde{\beta}_e \, G(z,z'), \qquad T_-=T=\tilde{\beta}_h \, G(z,z').
\end{equation}
Here we used the relation between the  wavevectors of exciton and carriers $\bm k^{e,h}=\bm q_\parallel m_{e,h}/(m_e+m_h)$ and introduced the constants
\begin{equation}
\tilde{\beta}_{e,h} = \beta_{e,h}{m_{e,h}\over m_e+m_h},
\end{equation}
where $m_h$ and $m_e$ are the heavy-hole mass in the QW plane
and the electron effective mass, respectively.
The function $G(z,z')$ describes a nonlocality of the QW exciton optical response:
\begin{equation}
G(z,z') = - 4\zeta {\Phi(z)\Phi^*(z') |d_\perp|^2\over (\hbar\omega_0 - \hbar\omega - i\Gamma)^2}.
\end{equation}
 
Solving the problem of light reflection from the QW in vicinity of the exciton resonance~\cite{EL_book}, we obtain the magnetogyrotropic correction to the Jones reflection matrix. For the incidence plane $(100)$ ($\bm q_\parallel \parallel [100]$) we get
\begin{align}
\label{dr_100}
 & \left[
\begin{array}{cc}
\Delta r_{s} & r_{sp}\\
r_{ps}&\Delta r_{p}
\end{array}
\right] = 4\zeta q_\parallel
 {i\Gamma_0\cos{\theta}\over (\hbar\omega_0 - \hbar\omega - i\Gamma)^2} \nonumber \\ 
& \times  \left[
\begin{array}{cc}
(\tilde{\beta}_e-\tilde{\beta}_h)B_x   &\tilde{\beta}_h\cos{\theta}B_y\\
\tilde{\beta}_h\cos{\theta}B_y&(\tilde{\beta}_e+\tilde{\beta}_h)\cos^2{\theta}B_x  
\end{array}
\right] .
\end{align}
Here $q_\parallel = (\omega/c)\sin{\theta_0}$, $\theta_0$ and $\theta$ are the light incidence angle and the angle of light propagation inside the sample, respectively,
and $\Gamma_0=2\pi\hbar|d_\perp|^2|\Lambda_0^2|q/\varepsilon_b$ is the $X_{hh}$ oscillator strength for normal incidence with $\varepsilon_b$ being the background dielectric constant of the QW material.

In the $\langle 110\rangle$ axes, $x'\parallel [1\bar{1}0]$, $y'\parallel [110]$, the BIA spin-orbit interaction~\eqref{H_SO} has the form
\begin{equation}
\label{H_SO_110}
{\cal H}_\text{SO} = \beta_e(\sigma_{x'}^e k_{y'}^e + \sigma_{y'}^e k_{x'}^e) + \beta_h(\sigma_{x'}^h k_{y'}^h + \sigma_{y'}^h k_{x'}^h).
\end{equation}
Calculation for the incidence plane $(110)$ when $\bm q_\parallel \parallel [110]$ yields
\begin{align}
\label{dr_110}
 & \left[
\begin{array}{cc}
\Delta r_{s} & r_{sp}\\
r_{ps}&\Delta r_{p}
\end{array}
\right] = 4\zeta q_\parallel
  {i\Gamma_0\cos{\theta}\over (\hbar\omega_0 - \hbar\omega - i\Gamma)^2} \nonumber \\ 
& \times  \left[
\begin{array}{cc}
(\tilde{\beta}_e+\tilde{\beta}_h)B_{y'}   &-\tilde{\beta}_h\cos{\theta}B_{x'}\\
-\tilde{\beta}_h\cos{\theta}B_{x'}&(\tilde{\beta}_e-\tilde{\beta}_h)\cos^2{\theta}B_{y'}  
\end{array}
\right] .
\end{align}
In this coordinate system, the reflection plane $(110)$ exists. The corresponding reflection keeps the combination $q_{x'}B_{y'}$ invariant and, hence, it can be present in $\Delta r_{s}$ and $\Delta r_{p}$. The combination $q_{x'}B_{x'}$ changes its sign under the reflection in the $(110)$ plane, and this allows for the corresponding term in $r_{sp}=r_{ps}$ because, at this reflection, the $s$ and $p$ components of the electric field are odd and even, respectively.

\section{Discussion}
\label{Disc}

The derived expressions demonstrate that the geometry $\bm q_\parallel \parallel \bm B_\parallel \parallel \langle 100 \rangle$ is suitable for observation of $\Delta r_{s}$ and $\Delta r_{p}$ caused by BIA, while in the geometry $\bm q_\parallel \parallel \bm B_\parallel \parallel \langle 110 \rangle$ the polarization conversion coefficient $r_{sp}=r_{ps}$ induced by BIA can be measured.
%
The experimental configuration used in the present work, Fig.~\ref{fig:Reflectance}, $\bm q_\parallel \parallel \bm B$, allows for maximal corrections to the reflection coefficients $\Delta r_s$ and $\Delta r_p$. In addition, these corrections have the resonant behavior $\propto (\omega_0-\omega-i\Gamma/\hbar)^{-2}$ at the $X_{hh}$ frequency. This is clearly seen in the experimental data, Figs.~\ref{fig:MSD_spectrum_CdTe} and~\ref{fig:Waterfall_GaAs}.

The derived Eqs.~\eqref{dr_100},~\eqref{dr_110} yield the following 
estimation for the magnetogyrotropic corrections: $\Delta r_{s,p} \sim (\beta q\Gamma_0/\Gamma^2) (a/l_B)^2$, where $a$ is the QW width and $l_B$ is the magnetic length. For $q=2.5\times 10^4$~cm$^{-1}$, $\beta=140$~meV~\AA~\cite{MD_MG_EL_PRB_2014}, $\Gamma=1$~meV, $\Gamma_0=0.1$~meV and $a=100$~\AA, we obtain $\Delta r_{s,p} \sim 10^{-3}B$~T~$^{-1}$. This value agrees in the order of magnitude  with the experimental data for both GaAs and CdTe based QWs, Figs.~\ref{fig:Amplitudes_CdTe} and~\ref{fig:Amplitudes_GaAs}.

We have also estimated other magnetogyrotropic contributions caused by BIA. Account for $\bm k$-odd terms in the bulk valence-band Hamiltonian as well as the interface inversion asymmetry terms also results in magnetogyrotropy due to admixture of the $\mp 1/2$ states to the $\pm 3/2$ states. However the corresponding contribution to the reflection has an order $\gamma_v q /(\Delta E_{lh}l_B^2)$ where $\gamma_v$ is the valence-band cubic in $k$ spin-orbit splitting constant and $\Delta E_{lh} \sim 10$~meV is the energy splitting between the size-quantized heavy- and light-hole levels. This value at $B=1$~T has an order of $10^{-5}$ which is two orders of magnitude smaller than the contributions~\eqref{dr_100},~\eqref{dr_110}.
Therefore we see that the BIA-induced spin-orbit splitting of the electron and hole states in QWs gives the dominant contribution to the magnetogyrotropic corrections to the reflection coefficients.

The ratio of the corrections for $p$ and $s$ polarized incident light according to Eq.~\eqref{dr_100} is given by
\begin{equation}
\left| {\Delta r_{p} \over \Delta r_{s}} \right| = \cos^2{\theta} \left| {\tilde{\beta}_e+\tilde{\beta}_h\over \tilde{\beta}_e-\tilde{\beta}_h}\right| .
\end{equation}
In the experiment, the signal for $p$ polarization is about twice stronger for both samples, see Figs.~\ref{fig:Amplitudes_CdTe} and~\ref{fig:Amplitudes_GaAs}. 
Since $\cos^2{\theta}\approx 1$ and $m_e \approx m_h$, the ratio 
agrees with the experiment at coinciding signs of ${\beta}_e$ and ${\beta}_h$ and at ${\beta}_h \approx 3{\beta}_e$.

\section{Conclusion}
\label{Concl}

From magnetoreflection experiments in vicinity of exciton resonances  we registered and analysed the magnetogyrotropic terms in the optical response of II-VI and II-V semiconductor QWs. We demonstrate that the $q_\parallel$- and $B_\parallel$-linear contribution to the reflection has an order of 0.1~\% in both QWs under study. 
The developed theory accounting for BIA spin-orbit splittings of electron and hole states in QWs agrees with the experimental findings.
Comparison of the  theory with experimental data allowed for determination of the ratio of the electron and heavy-hole BIA spin-splitting constants.

\acknowledgments 
We thank I.~A.~Akimov, A.~N.~Poddubny and E. L. Ivchenko for fruitful discussions.  
L.~V.~K. is supported by Russian Science Foundation (project 16-12-10503). 
L.~E.~G. thanks the Presidium of RAS 
 and the
Foundation for advancement of theoretical physics and mathematics ``BASIS''.

\end{document}